\documentstyle[twoside,fleqn,espcrc2]{article}

\input epsf
\newcommand\jhep[3]{J.\ High Energy Phys.\ {\bf #1} (#2) #3}
\newcommand\npb[3]{Nucl.\ Phys.\ {\bf B#1} (#2) #3}
\newcommand\plb[3]{Phys.\ Lett.\ {\bf B#1} (#2) #3}
\newcommand\prd[3]{Phys.\ Rev.\ {\bf D#1} (#2) #3}

\newcommand\prl[3]{Phys.\ Rev.\ Lett.\ {\bf #1} (#2) #3}

\newcommand\epjc[3]{Eur.\ Phys.\ J.\ {\bf C#1} (#2) #3}
\newcommand\ibid[3]{{\it ibid.\/} {\bf #1} (#2) #3}
\newcommand{\hepph}[1]{{\tt hep-ph/#1}}

\begin{document}

\begin{flushright}
SLAC-PUB-8266\\ 
September 1999\\[0.3cm]
hep-ph/9909564
\end{flushright}

\vspace{1.5cm}
\begin{center}
{\Large\bf\boldmath 
Model-independent information on $\gamma$ from $B^\pm\to\pi K$ 
Decays
\unboldmath}
\end{center}
 
\vspace{1.2cm}
\begin{center}
Matthias Neubert\footnote{Work supported by Department of Energy 
contract DE--AC03--76SF00515.}\\
{\sl Stanford Linear Accelerator Center, Stanford University\\
Stanford, California 94309, USA}\\[0.2cm]
and\\[0.2cm]
{\sl Newman Laboratory of Nuclear Studies, Cornell University\\
Ithaca, NY 14853, USA}
\end{center}
 
\vspace{1.3cm}
\begin{center}
{\bf Abstract:}\\[0.3cm]
\parbox{12cm}{ 
Measurements of the rates for the hadronic decays $B^\pm\to\pi K$
can be used to derive information on the weak phase $\gamma=
\mbox{arg}(V_{ub}^*)$ in a largely model-independent way. 
Hadronic uncertainties can be reduced to the level of 
nonfactorizable contributions to the decay amplitudes that are
power-suppressed in $\Lambda/m_b$ and, in addition, either 
violate SU(3) flavor symmetry or are doubly Cabibbo suppressed. 
Various strategies to obtain bounds on $\gamma$ and to extract 
its value with small theoretical uncertainty are described. The 
potential of $B^\pm\to\pi K$ decays for probing physics beyond 
the Standard Model is also discussed.}
\end{center}
 
\vspace{1.5cm}
\begin{center}
{\sl Invited talk presented at the\\
High-Energy Physics International Euroconference on Quantum
Chromo-Dynamics (QCD '99)\\[0.05cm]
Montpellier, France, 7--13 July 1999}
\end{center}
 
\newpage
\thispagestyle{empty}
\vbox{}\newpage
\setcounter{page}{1}

\title{Model-independent information on $\gamma$ from 
$B^\pm\to\pi K$ Decays}

\author{M. Neubert\address{
Stanford Linear Accelerator Center, Stanford University\\
Stanford, California 94309, USA\\
and\\
Newman Laboratory of Nuclear Studies, Cornell University\\
Ithaca, NY 14853, USA}}


\begin{abstract}
Measurements of the rates for the hadronic decays $B^\pm\to\pi K$
can be used to derive information on the weak phase $\gamma=
\mbox{arg}(V_{ub}^*)$ in a largely model-independent way. 
Hadronic uncertainties can be reduced to the level of 
nonfactorizable contributions to the decay amplitudes that are
power-suppressed in $\Lambda/m_b$ and, in addition, either 
violate SU(3) flavor symmetry or are doubly Cabibbo suppressed. 
Various strategies to obtain bounds on $\gamma$ and to extract 
its value with small theoretical uncertainty are described. The 
potential of $B^\pm\to\pi K$ decays for probing physics beyond 
the Standard Model is also discussed.
\end{abstract}

\maketitle

\section{INTRODUCTION}

The main objective of the $B$ factories is to explore in detail 
the physics of CP violation, to determine many of the flavor 
parameters of the electroweak theory, and to probe for possible 
effects of physics beyond the Standard Model. This 
will test the Cabibbo--Kobayashi--Maskawa (CKM) mechanism, which 
predicts that all CP violation results from a single complex 
phase in the quark mixing matrix. Facing the announcement of 
evidence for a CP asymmetry in the decays $B\to J/\psi\,K_S$ by 
the CDF Collaboration \cite{CDF}, the confirmation of direct CP 
violation in $K\to\pi\pi$ decays by the KTeV and NA48 
Collaborations \cite{KTeV,NA48}, and the  
successful start of the asymmetric $B$ factories at SLAC and KEK, 
the year 1999 has been an important step in achieving this goal.

The precise determination of the sides and angles of the 
``unitarity triangle'' $V_{ub}^* V_{ud}+V_{cb}^* V_{cd}+V_{tb}^* 
V_{td}=0$ plays a central role in the $B$-factory program 
\cite{BaBar}. With the standard phase conventions for the CKM 
matrix, only the two smallest elements in this relation, 
$V_{ub}^*$ and $V_{td}$, have nonvanishing imaginary parts (to 
an excellent approximation). In the Standard Model the angle 
$\beta=-\mbox{arg}(V_{td})$ can be determined in a theoretically 
clean way by measuring the time-dependent, mixing-induced CP 
asymmetry in the decays $B,\bar B\to J/\psi\,K_S$. The preliminary 
CDF result implies $\sin2\beta=0.79_{-0.44}^{+0.41}$ \cite{CDF}. 
The angle $\gamma=\mbox{arg}(V_{ub}^*)$, or equivalently the 
combination $\alpha=180^\circ-\beta-\gamma$, is much 
harder to determine \cite{BaBar}. Recently, there has been 
significant progress in the theoretical understanding of the 
hadronic decays $B\to\pi K$, and methods have been developed to 
extract information on $\gamma$ from rate measurements for these 
processes. Here we discuss the charged modes $B^\pm\to\pi K$, 
which are particularly clean from a theoretical perspective 
\cite{us,us2,me}. For applications involving the neutral 
decay modes the reader is referred to the literature 
\cite{FM,Robert}. 

In the Standard Model the main contributions to the decay 
amplitudes for the rare decays $B\to\pi K$ come from 
the penguin-induced flavor-changing neutral current (FCNC) 
transitions $\bar b\to\bar s q\bar q$, which exceed a small, 
Cabibbo-suppressed $\bar b\to\bar u u\bar s$ contribution 
from $W$-boson exchange. The weak phase $\gamma$ enters through
the interference of these two (``tree'' and ``penguin'') 
contributions. Because of a fortunate interplay of isospin, 
Fierz and flavor symmetries, the theoretical description of the 
charged modes $B^\pm\to\pi K$ is very clean despite the fact 
that these are exclusive nonleptonic decays \cite{us,us2,me}. 
Without any dynamical assumption, the hadronic uncertainties in 
the description of the interference terms relevant to the 
determination of $\gamma$ are of relative magnitude 
$O(\lambda^2)$ or $O(\epsilon_{\rm SU(3)}/N_c)$, where 
$\lambda=\sin\theta_C\approx 0.22$ is a measure of Cabibbo 
suppression, $\epsilon_{\rm SU(3)}\sim 20\%$ is the typical size 
of SU(3) flavor-symmetry breaking, and the factor $1/N_c$ 
indicates that the corresponding terms vanish in the factorization 
approximation. Factorizable SU(3) breaking can be accounted for in 
a straightforward way. 

Recently, the accuracy of this description has been further 
increased, because it has been shown that nonleptonic $B$ decays 
into two light mesons, such as $B\to\pi K$ and $B\to\pi\pi$, admit 
a heavy-quark expansion \cite{fact}. To leading order in 
$\Lambda/m_b$, and to all orders in perturbation theory, the decay 
amplitudes for these processes can be calculated from first 
principles, without recourse the phenomenological models. The QCD 
factorization theorem proved in \cite{fact} improves upon the 
phenomenological approach of ``generalized factorization'' 
\cite{Stech}, which emerges as the leading term in the  
heavy-quark limit. With the help of this theorem the irreducible 
theoretical uncertainty in the description of the $B^\pm\to\pi K$ 
decay amplitudes can be reduced by an extra factor of 
$O(\Lambda/m_b)$, rendering their analysis essentially model 
independent. 

As a consequence of this fact, and because they are dominated by 
(hadronic) FCNC transitions, the decays $B^\pm\to\pi K$ offer 
a sensitive probe to physics beyond the Standard Model 
\cite{me,Mati,CDK,anom,troja}, much in the same way as the 
``classical'' FCNC processes $B\to X_s\gamma$ or 
$B\to X_s\,l^+ l^-$. We will discuss how the bound on $\gamma$ 
and the extraction of $\gamma$ in the Standard Model could be 
affected by New Physics.

\section{\boldmath THEORY OF $B^\pm\to\pi K$ DECAYS\unboldmath}

The hadronic decays $B\to\pi K$ are mediated by a low-energy
effective weak Hamiltonian, whose operators allow for three
distinct classes of flavor topologies: QCD penguins, trees, and 
electroweak penguins. In the Standard Model the weak couplings 
associated with these topologies are known. From the measured 
branching ratios for the various $B\to\pi K$ decay modes it 
follows that QCD penguins dominate the decay amplitudes 
\cite{Digh}, whereas trees and electroweak penguins are 
subleading and of a similar strength \cite{oldDesh}. The 
theoretical description of the two charged modes 
$B^\pm\to\pi^\pm K^0$ and $B^\pm\to\pi^0 K^\pm$ exploits the fact 
that the amplitudes for these processes differ in a pure isospin 
amplitude $A_{3/2}$, defined as the matrix element of the 
isovector part of the effective Hamiltonian between a $B$ meson 
and the $\pi K$ isospin eigenstate with $I=\frac 32$. In the 
Standard Model the parameters of this amplitude are determined, 
up to an overall strong phase $\phi$, in the limit of SU(3) 
flavor symmetry \cite{us}. Using the QCD factorization theorem 
proved in \cite{fact}, the SU(3)-breaking corrections can be 
calculated in a model-independent way up to nonfactorizable 
terms that are power-suppressed in $\Lambda/m_b$ and vanish in 
the heavy-quark limit. 

A convenient parameterization of the decay amplitudes 
${\cal A}_{+0}\equiv{\cal A}(B^+\to\pi^+ K^0)$ and
${\cal A}_{0+}\equiv -\sqrt2\,{\cal A}(B^+\to\pi^0 K^+)$ is 
\cite{me}
\begin{eqnarray}\label{ampls}
   {\cal A}_{+0} &=& P\,(1-\varepsilon_a\,e^{i\gamma} e^{i\eta})
    \,, \\
   {\cal A}_{0+} &=& P \Big[ 1 - \varepsilon_a\,e^{i\gamma}
    e^{i\eta} - \varepsilon_{3/2}\,e^{i\phi}
    (e^{i\gamma} - \delta_{\rm EW}) \Big] \,,\nonumber
\end{eqnarray}
where $P$ is the dominant penguin amplitude defined as the sum 
of all terms in the $B^+\to\pi^+ K^0$ amplitude not proportional 
to $e^{i\gamma}$, $\eta$ and $\phi$ are strong phases, and 
$\varepsilon_a$, $\varepsilon_{3/2}$ and $\delta_{\rm EW}$ are 
real hadronic parameters. The weak phase $\gamma$ changes sign 
under a CP transformation, whereas all other parameters stay 
invariant. 

Let us discuss the various terms entering the decay amplitudes 
in detail. From a naive quark-diagram analysis one does not 
expect the $B^+\to\pi^+ K^0$ amplitude to receive a contribution 
from $\bar b\to\bar u u\bar s$ tree topologies; however, such a 
contribution can be induced through final-state rescattering or 
annihilation contributions \cite{Blok,Rob,Ge97,Ne97,Fa97,At97}. 
They are parameterized by $\varepsilon_a=O(\lambda^2)$. In the 
heavy-quark limit this parameter can be calculated  
and is found to be very small, $\varepsilon_a\approx-2\%$ 
\cite{newfact}. In the future, it will be possible to put upper 
and lower bounds on $\varepsilon_a$ by comparing the CP-averaged 
branching ratios for the decays $B^\pm\to\pi^\pm K^0$ and 
$B^\pm\to K^\pm\bar K^0$ \cite{Fa97}. Below we assume 
$|\varepsilon_a|\le 0.1$; however, our results will be almost 
insensitive to this assumption.

The terms proportional to $\varepsilon_{3/2}$ in (\ref{ampls}) 
parameterize the isospin amplitude $A_{3/2}$. The contribution 
proportional to $e^{i\gamma}$ comes from the tree process 
$\bar b\to\bar u u\bar s$, whereas the quantity $\delta_{\rm EW}$ 
describes the effects of electroweak penguins. The parameter 
$\varepsilon_{3/2}$ measures the relative strength of tree and 
QCD penguin contributions. Information about it can be derived by 
using SU(3) flavor symmetry to relate the tree contribution to 
the isospin amplitude $A_{3/2}$ to the corresponding contribution 
in the decay $B^+\to\pi^+\pi^0$. Since the final state 
$\pi^+\pi^0$ has isospin $I=2$ (because of Bose symmetry), the 
amplitude for this process does not receive any contribution from 
QCD penguins. Moreover, electroweak penguins in 
$\bar b\to\bar d q\bar q$ transitions are negligibly small. We 
define a related parameter $\bar\varepsilon_{3/2}$ by writing 
$\varepsilon_{3/2}=\bar\varepsilon_{3/2}
\sqrt{1-2\varepsilon_a\cos\eta\cos\gamma+\varepsilon_a^2}$, so
that the two quantities agree in the limit $\varepsilon_a\to 0$. 
In the SU(3) limit, this new parameter can be determined 
experimentally form the relation \cite{us}
\begin{equation}\label{eps}
   \bar\varepsilon_{3/2} = R_1
   \left|\frac{V_{us}}{V_{ud}}\right| \left[
   \frac{2\mbox{B}(B^\pm\to\pi^\pm\pi^0)}
        {\mbox{B}(B^\pm\to\pi^\pm K^0)} \right]^{1/2} .
\end{equation}
SU(3)-breaking corrections are described by the factor 
$R_1=1.22\pm 0.05$, which can be calculated in a 
model-independent way using the QCD factorization theorem for
nonleptonic decays \cite{newfact}. The quoted error is an 
estimate of the theoretical uncertainty due to uncontrollable 
corrections of $O(\frac{1}{N_c}\frac{m_s}{m_b})$. Using 
preliminary data reported by the CLEO Collaboration \cite{CLEO} 
to evaluate the ratio of CP-averaged branching ratios in 
(\ref{eps}) we obtain
\begin{equation}\label{epsval}
   \bar\varepsilon_{3/2} = 0.21\pm 0.06_{\rm exp}
   \pm 0.01_{\rm th} \,.
\end{equation}
With a better measurement of the branching ratios the uncertainty 
in $\bar\varepsilon_{3/2}$ will be reduced significantly.

Finally, the parameter
\begin{eqnarray}\label{delta}
   \delta_{\rm EW} &=& R_2\,
    \left| \frac{V_{cb}^* V_{cs}}{V_{ub}^* V_{us}} \right|\,
    \frac{\alpha}{8\pi}\,\frac{x_t}{\sin^2\!\theta_W}
    \left( 1 + \frac{3\ln x_t}{x_t-1} \right) \nonumber\\
   &=& (0.64\pm 0.09)\times\frac{0.085}{|V_{ub}/V_{cb}|} \,,
\end{eqnarray}
with $x_t=(m_t/m_W)^2$, describes the ratio of electroweak 
penguin and tree contributions to the isospin amplitude $A_{3/2}$. 
In the SU(3) limit it is calculable in terms of Standard Model 
parameters \cite{us,Fl96}. SU(3)-breaking corrections are 
accounted for by the quantity $R_2=0.92\pm 0.09$ \cite{me,newfact}. 
The error quoted in (\ref{delta}) also includes the uncertainty in 
the top-quark mass. 

Important observables in the study of the weak phase $\gamma$
are the ratio of the CP-averaged branching ratios in the two
$B^\pm\to\pi K$ decay modes,
\begin{equation}\label{Rst}
   R_* = \frac{\mbox{B}(B^\pm\to\pi^\pm K^0)}
              {2\mbox{B}(B^\pm\to\pi^0 K^\pm)} 
   = 0.75\pm 0.28 \,,
\end{equation}
and a particular combination of the direct CP asymmetries,
\begin{eqnarray}
   \widetilde A &=& \frac{A_{\rm CP}(B^\pm\to\pi^0 K^\pm)}{R_*}
    - A_{\rm CP}(B^\pm\to\pi^\pm K^0) \nonumber\\
   &=& - 0.52\pm 0.42 \,.
\end{eqnarray}
The experimental values of these quantities are derived from
preliminary data reported by the CLEO Collaboration 
\cite{CLEO}. The theoretical expressions for $R_*$ and 
$\widetilde A$ obtained using the parameterization in 
(\ref{ampls}) are
\begin{eqnarray}\label{expr}
   R_*^{-1} &=& 1 + 2\bar\varepsilon_{3/2}\cos\phi\,
    (\delta_{\rm EW}-\cos\gamma) \nonumber\\
   &+& \bar\varepsilon_{3/2}^2
    (1-2\delta_{\rm EW}\cos\gamma+\delta_{\rm EW}^2)
    + O(\bar\varepsilon_{3/2}\,\varepsilon_a) \,, \nonumber\\
   \widetilde A &=& 2\bar\varepsilon_{3/2} \sin\gamma \sin\phi
    + O(\bar\varepsilon_{3/2}\,\varepsilon_a) \,.
\end{eqnarray}
Note that the rescattering effects described by $\varepsilon_a$ 
are suppressed by a factor of $\bar\varepsilon_{3/2}$ and thus 
reduced to the percent level. Explicit expressions for these
contributions can be found in \cite{me}.

\section{\boldmath LOWER BOUND ON $\gamma$ AND CONSTRAINT IN THE 
$(\bar\rho,\bar\eta)$ PLANE\unboldmath}

There are several strategies for exploiting the above relations.
First, from a measurement of the ratio $R_*$ alone a bound on
$\cos\gamma$ can be derived, which implies a nontrivial 
constraint on the Wolfenstein parameters $\bar\rho$ and 
$\bar\eta$ defining the apex of the unitarity triangle \cite{us}. 
Only CP-averaged branching ratios are needed for this purpose. 
Varying the strong phases $\phi$ and $\eta$ independently we 
first obtain an upper bound on the inverse of $R_*$. Keeping 
terms of linear order in $\varepsilon_a$, we find \cite{me}
\begin{eqnarray}\label{Rbound}
   R_*^{-1} &\le& \left( 1 + \bar\varepsilon_{3/2}\,
    |\delta_{\rm EW}-\cos\gamma| \right)^2
    + \bar\varepsilon_{3/2}^2\sin^2\!\gamma \nonumber\\
   &+& 2\bar\varepsilon_{3/2}|\varepsilon_a|\sin^2\!\gamma \,.
\end{eqnarray}
Provided $R_*$ is significantly smaller than 1, this bound 
implies an exclusion region for $\cos\gamma$, which becomes 
larger the smaller the values of $R_*$ and 
$\bar\varepsilon_{3/2}$ are. It is convenient to consider 
instead of $R_*$ the related quantity \cite{troja}
\begin{equation}\label{Rval}
   X_R = \frac{\sqrt{R_*^{-1}}-1}{\bar\varepsilon_{3/2}} 
   = 0.72\pm 0.98_{\rm exp}\pm 0.03_{\rm th} \,.
\end{equation}
Because of the theoretical factor $R_1$ entering the definition 
of $\bar\varepsilon_{3/2}$ in (\ref{eps}) this is, strictly 
speaking, not an observable. However, the theoretical uncertainty 
in $X_R$ is so much smaller than the present experimental 
error that it is justified to treat this quantity as an 
observable. The advantage of presenting our results in terms of 
$X_R$ rather than $R_*$ is that the leading dependence on 
$\bar\varepsilon_{3/2}$ cancels out, leading to the simple bound 
$|X_R|\le|\delta_{\rm EW}-\cos\gamma|+O(\bar\varepsilon_{3/2},
\varepsilon_a)$.

\begin{figure}
\epsfxsize=7cm
\centerline{\epsffile{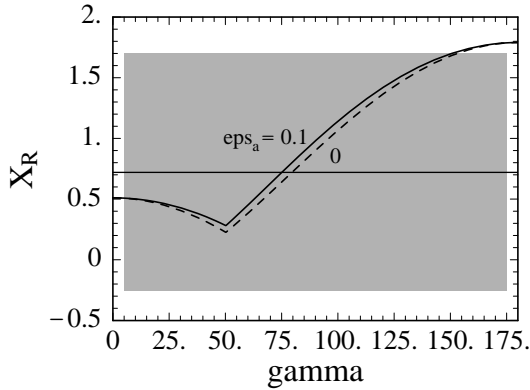}}
\vspace{-0.75cm}
\caption{\label{fig:Rbound}
Theoretical upper bound on the ratio $X_R$ versus 
$|\gamma|$ for $\varepsilon_a=0.1$ (solid) and $\varepsilon_a=0$ 
(dashed). The horizontal line and band show the current 
experimental value with its $1\sigma$ variation.}
\end{figure}

In Figure~\ref{fig:Rbound} we show the upper bound on $X_R$
as a function of $|\gamma|$, obtained by varying the input 
parameters in the intervals $0.15\le\bar\varepsilon_{3/2}\le 
0.27$ and $0.49\le\delta_{\rm EW}\le 0.79$ (corresponding to 
using $|V_{ub}/V_{cb}|=0.085\pm 0.015$ in (\ref{delta})). 
Note that the effect of the rescattering contribution 
parameterized by $\varepsilon_a$ is very small. The gray band 
shows the current value of $X_R$, which clearly has too large 
an error to provide any useful information on 
$\gamma$.\footnote{Unfortunately, the $2\sigma$ deviation from 
1 indicated by the first preliminary CLEO result has not been 
confirmed by the present data.} 
The situation may change, however, once a more precise 
measurement of $X_R$ will become available. For instance, if 
the current central value $X_R=0.72$ were confirmed, it would 
imply the bound $|\gamma|>75^\circ$, which would mark a 
significant improvement over the limit $|\gamma|>37^\circ$ 
obtained from the global analysis of the unitarity triangle 
including information from $K$--$\bar K$ mixing \cite{BaBar}. 

\begin{figure}
\epsfxsize=8cm
\centerline{\epsffile{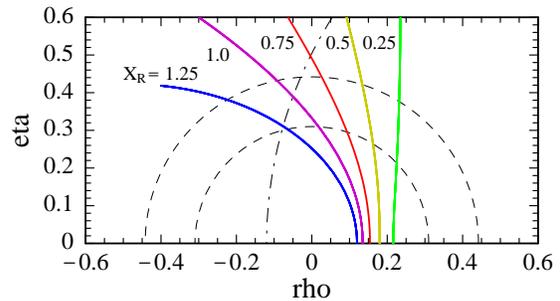}}
\vspace{-0.75cm}
\caption{\label{fig:CKMbound}
Theoretical constraint on the Wolfenstein parameters 
$(\bar\rho,\bar\eta)$ implied by a measurement of the ratio 
$X_R$ in $B^\pm\to\pi K$ decays (solid lines), semileptonic $B$ 
decays (dashed circles), and $B_s$--$\bar B_s$ mixing 
(dashed-dotted line).}
\end{figure}

So far, as in previous work, we have used the inequality 
(\ref{Rbound}) to derive a lower bound on $|\gamma|$. However, a 
large part of the uncertainty in the value of $\delta_{\rm EW}$, 
and thus in the resulting bound on $|\gamma|$, comes from the 
present large error on $|V_{ub}|$. Since this is not a hadronic 
uncertainty, it is more appropriate to separate it and turn 
(\ref{Rbound}) into a constraint on the Wolfenstein parameters 
$\bar\rho$ and $\bar\eta$. To this end, we use that $\cos\gamma
=\bar\rho/\sqrt{\bar\rho^2+\bar\eta^2}$ by definition, and 
$\delta_{\rm EW}=(0.24\pm 0.03)/\sqrt{\bar\rho^2+\bar\eta^2}$ 
from (\ref{delta}). The solid lines in Figure~\ref{fig:CKMbound} 
show the resulting constraint in the $(\bar\rho,\bar\eta)$ plane 
obtained for the representative values $X_R=0.25$, 0.5, 0.75, 1.0,
1.25 (from right to left), which for $\bar\varepsilon_{3/2}=0.21$
would correspond to $R_*=0.90$, 0.82, 0.75, 0.68, 0.63, 
respectively. Values to the right of these lines are excluded. For 
comparison, the dashed circles show the constraint arising from the 
measurement of the ratio $|V_{ub}/V_{cb}|=0.085\pm 0.015$ in 
semileptonic $B$ decays, and the dashed-dotted line shows the bound 
implied by the present experimental limit on the mass difference 
$\Delta m_s$ in the $B_s$ system \cite{BaBar}. Values to the left 
of this line are excluded. It is evident from the figure that the 
bound resulting from a measurement of the ratio $X_R$ in 
$B^\pm\to\pi K$ decays may be very nontrivial and, in particular, 
may eliminate the possibility that $\gamma=0$. The combination of 
this bound with information from semileptonic decays and 
$B_s$--$\bar B_s$ mixing alone would then determine the 
Wolfenstein parameters $\bar\rho$ and $\bar\eta$ within narrow 
ranges,\footnote{An observation of CP violation, such as the 
measurement of $\epsilon_K$ in $K$--$\bar K$ mixing or 
$\sin2\beta$ in $B\to J/\psi\,K_S$ decays, is however needed to 
fix the sign of $\bar\eta$.}
and in the context of the CKM model would prove the existence of 
direct CP violation in $B$ decays.

\section{\boldmath EXTRACTION OF $\gamma$\unboldmath}
\label{sec:determ}

Ultimately, the goal is of course not only to derive a bound on 
$\gamma$ but to determine this parameter directly from the 
data. This requires to fix the strong phase $\phi$ in 
(\ref{expr}), which can be done either through the measurement 
of a CP asymmetry or with the help of theory. A strategy for an 
experimental determination of $\gamma$ from $B^\pm\to\pi K$ 
decays has been suggested in \cite{us2}. It generalizes a 
method proposed by Gronau, Rosner and London \cite{GRL} to 
include the effects of electroweak penguins. The approach has 
later been refined to account for rescattering contributions 
to the $B^\pm\to\pi^\pm K^0$ decay amplitudes \cite{me}. Before
discussing this method, we will first illustrate an easier 
strategy for a theory-guided determination of $\gamma$ based on
the QCD factorization theorem derived in \cite{fact}. This
method does not require any measurement of a CP asymmetry.

\subsection{Theory-guided determination}

In the previous section the theoretical predictions for the 
nonleptonic $B\to\pi K$ decay amplitudes obtained using the QCD 
factorization theorem were used in a minimalistic way, i.e., 
only to calculate the size of the SU(3)-breaking effects 
parameterized by $R_1$ and $R_2$. The resulting bound on 
$\gamma$ and the corresponding constraint in the 
$(\bar\rho,\bar\eta)$ plane are therefore theoretically very 
clean. However, they are only useful if the value of $X_R$ is 
found to be larger than about 0.5 (see 
Figure~\ref{fig:Rbound}), in which case values of $|\gamma|$ 
below $65^\circ$ are excluded. If it would turn out that 
$X_R<0.5$, then it is in principle possible to satisfy the 
inequality (\ref{Rbound}) also for small values of $\gamma$, 
however, at the price of having a very large value of the 
strong phase, $\phi\approx 180^\circ$. But this possibility can 
be discarded based on the model-independent prediction that 
\cite{fact}
\begin{equation}\label{phiest}
   \phi = O[\alpha_s(m_b),\Lambda/m_b] \,.
\end{equation}
In fact, a direct calculation of this phase to leading power
in $\Lambda/m_b$ yields $\phi\approx-11^\circ$ \cite{newfact}.
Using the fact that $\phi$ is parametrically small, we can 
exploit a measurement of the ratio $X_R$ to obtain a 
{\em determination\/} of $|\gamma|$ -- corresponding to an 
allowed region in the $(\bar\rho,\bar\eta)$ plane -- rather 
than just a bound. This determination is unique up to a sign. 
Note that for small values of $\phi$ the impact of the strong 
phase in the expression for $R_*$ in (\ref{expr}) is a 
second-order effect, since $\cos\phi\approx 1-\phi^2/2$. As 
long as $|\phi|\ll\sqrt{2\Delta\bar\varepsilon_{3/2}/
\bar\varepsilon_{3/2}}$, the uncertainty in the value of 
$\cos\phi$ has a much smaller effect than the uncertainty in 
$\bar\varepsilon_{3/2}$. With the present value of 
$\bar\varepsilon_{3/2}$, this is the case as long as 
$|\phi|\ll 43^\circ$. We believe it is a safe assumption to take 
$|\phi|<25^\circ$ (i.e., more than twice the value obtained 
to leading order in $\Lambda/m_b$), so that $\cos\phi>0.9$. 

\begin{figure}
\epsfxsize=8cm
\centerline{\epsffile{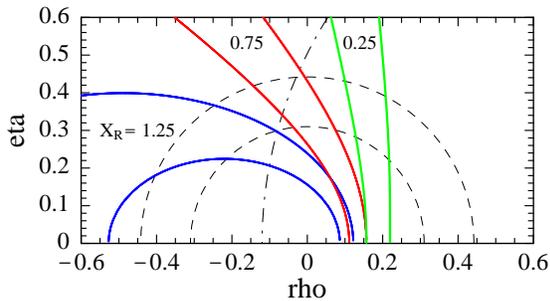}}
\vspace{-0.75cm}
\caption{\label{fig:CKMfit}
Allowed regions in the $(\bar\rho,\bar\eta)$ plane for fixed 
values of $X_R$, obtained by varying all theoretical parameters 
inside their respective ranges of uncertainty, as specified in 
the text. The sign of $\bar\eta$ is not determined.} 
\end{figure}

Solving the equation for $R_*$ in (\ref{expr}) for $\cos\gamma$,
and including the corrections of $O(\varepsilon_a)$, we find 
\begin{eqnarray}\label{gamth}
   \cos\gamma &=& \delta_{\rm EW}
    - \frac{X_R + \frac12\bar\varepsilon_{3/2}
            (X_R^2-1+\delta_{\rm EW}^2)}
           {\cos\phi+\bar\varepsilon_{3/2}\delta_{\rm EW}}
    \nonumber\\
   &+& \frac{\varepsilon_a\cos\eta\sin^2\!\gamma}
            {\cos\phi+\bar\varepsilon_{3/2}\delta_{\rm EW}}
    \,,
\end{eqnarray}
where we have set $\cos\phi=1$ in the $O(\varepsilon_a)$ term.
Using the QCD factorization theorem one finds that 
$\varepsilon_a\cos\eta\approx -0.02$ in the heavy-quark limit 
\cite{newfact}, and we assign a 100\% uncertainty to this 
estimate. In evaluating the result (\ref{gamth}) we scan 
the parameters in the ranges $0.15\le\bar\varepsilon_{3/2}\le
0.27$, $0.55\le\delta_{\rm EW}\le 0.73$, $-25^\circ\le\phi\le
25^\circ$, and $-0.04\le\varepsilon_a\cos\eta\sin^2\!\gamma
\le 0$. Figure~\ref{fig:CKMfit} shows the allowed regions 
in the $(\bar\rho,\bar\eta)$ plane for the representative 
values $X_R=0.25$, 0.75, and 1.25 (from right to left). We 
stress that with this method a useful constraint on the 
Wolfenstein parameters is obtained for {\em any\/} value of 
$X_R$.

\subsection{Model-independent determination}

\begin{figure}
\epsfxsize=6.1cm
\centerline{\epsffile{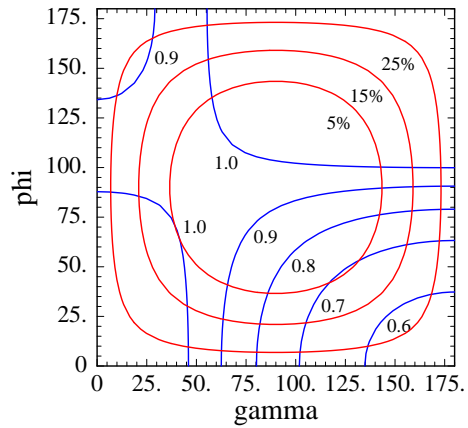}}
\vspace{-0.75cm}
\caption{\label{fig:contours}
Contours of constant $R_*$ (``hyperbolas'') and constant 
$|\widetilde A|$ (``circles'') in the $(|\gamma|,|\phi|)$ plane. 
The sign of the asymmetry $\widetilde A$ determines the sign of 
the product $\sin\gamma\sin\phi$. The contours for $R_*$ refer to 
values from 0.6 to 1.0 in steps of 0.1, those for the asymmetry 
correspond to 5\%, 15\%, and 25\%, as indicated.} 
\end{figure}

It is important that, once more precise data on $B^\pm\to\pi K$
decays will become available, it will be possible to test the
theoretical prediction of a small strong phase $\phi$
experimentally. To this end, one must determine the CP
asymmetry $\widetilde A$ in addition to the ratio $R_*$. From 
(\ref{expr}) it follows that for fixed values of 
$\bar\varepsilon_{3/2}$ and $\delta_{\rm EW}$ the quantities $R_*$ 
and $\widetilde A$ define contours in the $(\gamma,\phi)$ plane, 
whose intersections determine the two phases up to possible 
discrete ambiguities \cite{us2,me}. Figure~\ref{fig:contours} shows 
these contours for some representative values, assuming 
$\bar\varepsilon_{3/2}=0.21$, $\delta_{\rm EW}=0.64$, and 
$\varepsilon_a=0$. In practice, including the uncertainties in the 
values of these parameters changes the contour lines into contour 
bands. Typically, the spread of the bands induces an error in
the determination of $\gamma$ of about $10^\circ$ 
\cite{me}.\footnote{A precise determination of this error requires 
knowledge of the actual values of the observables. Gronau and 
Pirjol \protect\cite{Pirj} find a larger error for the special 
case where the product $|\sin\gamma\sin\phi|$ is very close to 1, 
which however is highly disfavored because of the expected 
smallness of the strong phase $\phi$.} 
In the most general case there are up to eight discrete solutions 
for the two phases, four of which are related to the other four 
by the sign change $(\gamma,\phi)\to(-\gamma,-\phi)$. However, for 
typical values of $R_*$ it turns out that often only four 
solutions exist, two of which are related to the other two by a 
change of signs. The theoretical prediction that $\phi$ is small 
implies that solutions should exist where the contours intersect 
close to the lower portion in the plot. Other solutions with large 
$\phi$ are strongly disfavored theoretically. Moreover, according 
to (\ref{expr}) the sign of the CP asymmetry $\widetilde A$ fixes 
the relative sign between the two phases $\gamma$ and $\phi$. If 
we trust the theoretical prediction that $\phi$ is negative 
\cite{newfact}, it follows that in most cases there remains only a 
unique solution for $\gamma$, i.e., {\em the CP-violating phase 
$\gamma$ can be determined without any discrete ambiguity}. 

As an example, consider the hypothetical case where $R_*=0.8$ 
and $\widetilde A=-15\%$. Figure~\ref{fig:contours} then allows 
the four solutions where $(\gamma,\phi)\approx
(\pm 82^\circ,\mp 21^\circ)$ or $(\pm 158^\circ,\mp 78^\circ)$. 
The second pair of solutions is strongly disfavored because of 
the large values of the strong phase $\phi$. From the first 
pair of solutions, the one with $\phi\approx -21^\circ$ is 
closest to our theoretical expectation that $\phi\approx 
-11^\circ$, hence leaving $\gamma\approx 82^\circ$ as the unique 
solution.

\section{SEARCH FOR NEW PHYSICS}

In the presence of New Physics the theoretical description of 
$B^\pm\to\pi K$ decays becomes more complicated. In particular, 
new CP-violating contributions to the decay amplitudes may be 
induced. A detailed analysis has been presented in \cite{troja}. 
A convenient and completely general parameterization of the two 
amplitudes in (\ref{ampls}) is obtained by replacing
\begin{eqnarray}\label{replace}
   P &\to& P' \,, \qquad
    \varepsilon_a\,e^{i\gamma} e^{i\eta} \to
    i\rho\,e^{i\phi_\rho} \,, \nonumber\\
   \delta_{\rm EW} &\to& a\,e^{i\phi_a} + ib\,e^{i\phi_b} \,,
\end{eqnarray}
where $\rho$, $a$, $b$ are real hadronic parameters, and 
$\phi_\rho$, $\phi_a$, $\phi_b$ are strong phases. The terms 
$i\rho$ and $ib$ change sign under a CP transformation. New Physics
effects parameterized by $P'$ and $\rho$ are isospin conserving,
while those described by $a$ and $b$ violate isospin. Note that the
parameter $P'$ cancels in all ratios of branching ratios and thus
does not affect the quantities $R_*$ and $X_R$ as well as all
CP asymmetries. Because the ratio $R_*$ in (\ref{Rst}) would be 1 
in the isospin limit, it is particularly sensitive to 
isospin-violating New Physics contributions. The isospin-conserving 
effects parameterized by $\rho$ enter only through interference 
with the isospin-violating terms proportional to 
$\varepsilon_{3/2}$ in (\ref{ampls}) and hence are suppressed. 

New Physics can affect the bound on $\gamma$ derived from 
(\ref{Rbound}) as well as the value of $\gamma$ extracted using
the strategies discussed in the previous section. We will discuss
these two possibilities in turn.

\subsection{\boldmath Effects on the bound on $\gamma$\unboldmath}

The upper bound on $R_*^{-1}$ in (\ref{Rbound}) and the 
corresponding bound on $X_R$ shown in Figure~\ref{fig:Rbound} are 
model-independent results valid in the Standard Model. Note that 
the extremal value of $R_*^{-1}$ is such that 
$|X_R|\le(1+\delta_{\rm EW})$ irrespective of $\gamma$. A value of 
$|X_R|$ exceeding this bound would be a clear signal for New 
Physics \cite{me,Mati,troja}. 

Consider first the case where New Physics may induce arbitrary 
CP-violating contributions to the $B\to\pi K$ decay amplitudes, 
while preserving isospin symmetry. Then the only change with respect 
to the Standard Model is that the parameter $\rho$ may no longer be 
as small as $O(\varepsilon_a)$. Varying the strong phases $\phi$ and 
$\phi_\rho$ independently, and allowing for an {\em arbitrarily 
large\/} New Physics contribution to $\rho$, one can derive the 
bound \cite{troja} 
\begin{equation}\label{phiarb}
   |X_R| \le \sqrt{1 - 2\delta_{\rm EW}\cos\gamma 
   + \delta_{\rm EW}^2} \le 1+\delta_{\rm EW} \,.
\end{equation}
Note that the extremal value is the same as in the Standard Model, 
i.e., isospin-conserving New Physics effects cannot lead to a value 
of $|X_R|$ exceeding $1+\delta_{\rm EW}$. For intermediate values 
of $\gamma$ between $25^\circ$ and $125^\circ$ the Standard Model
bound on $X_R$ is weakened. But even for large values $\rho=O(1)$, 
corresponding to a significant New Physics contribution to the 
decay amplitudes, the effects are small.

If both isospin-violating and isospin-conserving New Physics 
effects are present and involve new CP-violating phases, the 
analysis becomes more complicated. Still, it is possible to derive 
model-independent bounds on $X_R$. Allowing for arbitrary values 
of $\rho$ and all strong phases, one obtains \cite{troja}
\begin{eqnarray}\label{abbound}
   |X_R| &\le& \sqrt{(|a|+|\cos\gamma|)^2 + (|b|+|\sin\gamma|)^2}
    \nonumber\\
   &\le& 1 + \sqrt{a^2 + b^2}
    \le \frac{2}{\bar\varepsilon_{3/2}} + X_R \,,
\end{eqnarray}
where the last inequality is relevant only in cases where
$\sqrt{a^2 + b^2}\gg 1$. The important point to note is that with
isospin-violating New Physics contributions the value of $|X_R|$
can exceed the upper bound in the Standard Model by a potentially 
large amount. For instance, if $\sqrt{a^2+b^2}$ is twice as large
as in the Standard Model, corresponding to a New Physics 
contribution to the decay amplitudes of only 10--15\%, then $|X_R|$ 
could be as large as 2.6 as compared with the maximal value 1.8 
allowed in the Standard Model. Also, in the most general case where 
$b$ and $\rho$ are nonzero, the maximal value $|X_R|$ can take 
is no longer restricted to occur at the endpoints $\gamma=0^\circ$ 
or $180^\circ$, which are disfavored by the global analysis of the 
unitarity triangle \cite{BaBar}. Rather, $|X_R|$ would take its 
maximal value if $|\tan\gamma|=|\rho|=|b/a|$.

The present experimental value of $X_R$ in (\ref{Rval}) has too 
large an error to determine whether there is any deviation from 
the Standard Model. If $X_R$ turns out to be larger than 1 (i.e., 
one third of a standard deviation above its current central 
value), then an interpretation of this result in the Standard 
Model would require a large value $|\gamma|>91^\circ$ (see 
Figure~\ref{fig:Rbound}), which may be difficult to accommodate. 
This may be taken as evidence for New Physics. If $X_R>1.3$, one 
could go a step further and conclude that the New Physics must 
necessarily violate isospin \cite{troja}. 

\subsection{\boldmath Effects on the determination of $\gamma$
\unboldmath}

A value of the observable $R_*$ violating the Standard Model bound 
(\ref{Rbound}) would be an exciting hint for New Physics. However, 
even if a more precise measurement will give a value that is 
consistent with the Standard Model bound, $B^\pm\to\pi K$ decays 
provide an excellent testing ground for physics beyond the 
Standard Model. This is so because New Physics may still cause a 
significant shift in the value of $\gamma$ extracted from 
$B^\pm\to\pi K$ decays using the strategies discussed in 
Section~\ref{sec:determ}. This may lead to inconsistencies when 
this value is compared with other determinations of $\gamma$. 

A global fit of the unitarity triangle combining information from 
semileptonic $B$ decays, $B$--$\bar B$ mixing, CP violation in the 
kaon system, and mixing-induced CP violation in $B\to J/\psi\,K_S$ 
decays provides information on $\gamma$, which in a few years will 
determine its value within a rather narrow range \cite{BaBar}. 
Such an indirect determination could be complemented by direct 
measurements of $\gamma$ using, e.g., $B\to D K$ decays, or using 
the triangle relation $\gamma=180^\circ-\alpha-\beta$ combined with 
a measurement of $\alpha$ in $B\to\pi\pi$ or $B\to\pi\rho$ decays. 
We will assume that a discrepancy of more than $25^\circ$ between 
the ``true'' $\gamma=\mbox{arg}(V_{ub}^*)$ and the value 
$\gamma_{\pi K}$ extracted in $B^\pm\to\pi K$ decays will be 
observable after a few years of operation at the $B$ factories. This 
will set the benchmark for sensitivity to New Physics effects.

\begin{figure}
\epsfxsize=7cm
\centerline{\epsffile{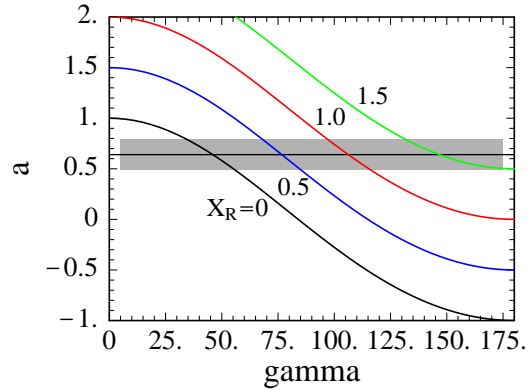}}
\vspace{-0.75cm}
\caption{\label{fig:shift}
Contours of constant $X_R$ versus $\gamma$ and the parameter $a$, 
assuming $\gamma>0$. The horizontal band shows the value of $a$ 
in the Standard Model.}
\end{figure}

In order to illustrate how big an effect New Physics could have 
on the value of $\gamma$ we consider the simplest case where there 
are no new CP-violating couplings. Then all New Physics 
contributions in (\ref{replace}) are parameterized by the single 
parameter $a\equiv\delta_{\rm EW}+a_{\rm NP}$. A more general 
discussion can be found in \cite{troja}. We also assume for 
simplicity that the strong phase $\phi$ is small, as suggested by 
(\ref{phiest}). In this case the difference between the value 
$\gamma_{\pi K}$ extracted from $B^\pm\to\pi K$ decays and the 
``true'' value of $\gamma$ is to a good approximation given by
\begin{equation}
   \cos\gamma_{\pi K} \simeq \cos\gamma - a_{\rm NP} \,.
\end{equation}
In Figure~\ref{fig:shift} we show contours of constant $X_R$ 
versus $\gamma$ and $a$, assuming without loss of generality that 
$\gamma>0$. Obviously, even a moderate New Physics contribution to 
the parameter $a$ can induce a large shift in $\gamma$. Note that 
the present central value of $X_R\approx 0.7$ is such that values 
of $a$ less than the Standard Model result $a\approx 0.64$ are 
disfavored, since they would require values of $\gamma$ exceeding 
$100^\circ$, in conflict with the global analysis of the unitarity 
triangle \cite{BaBar}.

\subsection{Survey of New Physics models}

In \cite{troja}, we have explored how physics beyond the Standard 
Model could affect purely hadronic FCNC transitions of the type 
$\bar b\to\bar s q\bar q$ focusing, in particular, on isospin 
violation. Unlike in the Standard Model, where isospin-violating 
effects in these processes are strongly suppressed by electroweak 
gauge couplings or small CKM matrix elements, in many New Physics 
scenarios these effects are not parametrically suppressed relative 
to isospin-conserving FCNC processes. In the language of effective 
weak Hamiltonians this implies that the Wilson coefficients of 
QCD and electroweak penguin operators are of a similar magnitude. 
For a large class of New Physics models we found that the 
coefficients of the electroweak penguin operators are, in fact, 
due to ``trojan'' penguins, which are neither related to penguin 
diagrams nor of electroweak origin. 

Specifically, we have considered: (a) models with tree-level FCNC 
couplings of the $Z$ boson, extended gauge models with an extra 
$Z'$ boson, supersymmetric models with broken R-parity; (b) 
supersymmetric models with R-parity conservation; (c) 
two-Higgs--doublet models, and models with anomalous gauge-boson 
couplings. Some of these models have also been investigated in 
\cite{CDK,anom}. In case (a), the resulting electroweak penguin 
coefficients can be much larger than in the Standard Model because 
they are due to tree-level processes. In case (b), these 
coefficients can compete with the ones of the Standard Model  
because they arise from strong-interaction box diagrams, which 
scale relative to the Standard Model like
$(\alpha_s/\alpha)(m_W^2/m_{\rm SUSY}^2)$. In models (c), on 
the other hand, isospin-violating New Physics effects are not
parametrically enhanced and are generally smaller than in the 
Standard Model.

\begin{table}
\caption{\label{tab:1}
Maximal contributions to $a_{\rm NP}$ in extensions of the 
Standard Model. Entries marked with a ``$^*$'' are upper bounds 
derived using (\protect\ref{abbound}). For the case of 
supersymmetric models with R-parity the first (second) row 
corresponds to maximal right-handed (left-handed) 
strange--bottom squark mixing. For the two-Higgs--doublet models
we take $m_{H^+}>100$\,GeV and $\tan\beta>1$.}
\vspace{0.2cm}
\begin{tabular}{lcc}
\hline
Model & $|a_{\rm NP}|$ & $|\gamma_{\pi K}-\gamma|$ \\
\hline
FCNC $Z$ exchange & 2.0 & $180^\circ$ \\
extra $Z'$ boson & 14$^*$ & $180^\circ$ \\
SUSY without R-parity & 14$^*$ & $180^\circ$ \\
\hline
SUSY with R-parity & 0.4 & $25^\circ$ \\
                   & 1.3 & $180^\circ$ \\
\hline
2HDM & 0.15 & $10^\circ$ \\
anom.\ gauge-boson coupl.\ & 0.3 & $20^\circ$ \\
\hline
\end{tabular}
\end{table}

For each New Physics model we have explored which region of 
parameter space can be probed by the $B^\pm\to\pi K$ observables, 
and how big a departure from the Standard Model predictions one 
can expect under realistic circumstances, taking into account all 
constraints on the model parameters implied by other processes. 
Table~\ref{tab:1} summarizes our estimates of the maximal 
isospin-violating contributions to the decay amplitudes, as 
parameterized by $|a_{\rm NP}|$. They are the potentially most 
important source of New Physics effects in $B^\pm\to\pi K$ 
decays. For comparison, we recall that in the Standard Model 
$a\approx 0.64$. Also shown are the corresponding maximal values 
of the difference $|\gamma_{\pi K}-\gamma|$. As noted above, in 
models with tree-level FCNC couplings New Physics effects can be 
dramatic, whereas in supersymmetric models with R-parity 
conservation isospin-violating loop effects can be competitive 
with the Standard Model. In the case of supersymmetric models 
with R-parity violation the bound (\ref{abbound}) implies 
interesting limits on certain combinations of the trilinear 
couplings $\lambda_{ijk}'$ and $\lambda_{ijk}''$, which are 
discussed in \cite{troja}.

\section{CONCLUSIONS}

Measurements of the rates for the rare hadronic decays 
$B^\pm\to\pi K$ provide interesting information on the weak 
phase $\gamma$ and on the Wolfenstein parameters $\bar\rho$ and 
$\bar\eta$. Using isospin, Fierz and flavor symmetries together 
with the fact that nonleptonic $B$ decays into two light mesons 
admit a heavy-quark expansion, a largely model-independent 
description of these decays is obtained despite the fact that 
they are exclusive nonleptonic processes. In the future, a precise 
measurement of the $B^\pm\to\pi K$ decay amplitudes will provide 
an extraction of $\gamma$ with a theoretical uncertainty of about 
$10^\circ$, and at the same time will allow for sensitive tests of 
physics beyond the Standard Model.

\subsection*{Acknowledgements}

It is a pleasure to thank the SLAC Theory Group for the warm
hospitality extended to me during the past year. I am grateful 
to Martin Beneke, Gerhard Buchalla, Yuval Grossman, Alex Kagan, 
Jon Rosner and Chris Sachrajda for collaboration on parts of the 
work reported here. This research was supported by the Department 
of Energy under contract DE--AC03--76SF00515.

\end{document}